\documentclass[conference]{IEEEtran}
\usepackage[letterpaper, left=0.7in, right=0.7in, bottom=0.95in, top=0.7in]{geometry}

\usepackage{algorithm}
\usepackage{algorithmic}
\usepackage[font=small]{caption}
\usepackage[labelsep=period]{caption}
\usepackage{float}
\usepackage[latin1]{inputenc}
\usepackage{graphicx}
\usepackage[cmex10]{amsmath}
\usepackage{cite}
\usepackage{amsfonts}
\usepackage{amsmath}
\usepackage{amsthm}
\usepackage{mathrsfs,dsfont}
\usepackage{amssymb,mathrsfs}
\usepackage[mathscr]{euscript}
\usepackage{epstopdf} 
\usepackage{color}
\usepackage{pgfplots}
\pgfplotsset{compat=newest}
\usetikzlibrary{plotmarks}
\usepackage{grffile}
\usepackage{amsmath}
\usepackage{caption}
\usepackage{subcaption}
\usepackage{balance}

\newtheorem{remark}{Remark}

\IEEEoverridecommandlockouts

\def \hu{\mathrm{h_U}}
\def \mcp{\mathrm{m_{cp}}}
\def \fcp{\mathrm{f_{cp}}}
\def \mUAV{\mathrm{m_{U}}}
\def \lamc{\lambda_\mathrm{C}}
\def \lammec{\lambda_\mathrm{M}}
\def \fc{\mathrm{f_c}}
\def \hbuild{\mathrm{h_b}}
\def \hbs{\mathrm{h_B}}
\def \Vc{\mathrm{V}}
\def \Ptx{\mathrm{P_{tx}}}
\def \rcomm{\mathcal{R}_\mathrm{cm}}
\def \rcomp{\mathcal{R}_\mathrm{cp}}
\def \rtot{\mathcal{R}_\mathrm{tot}}
\def \th{\mathrm{T_H}}
\def \tmr{\mathrm{T_{MR}}}
\def \eh{\mathcal{E}_\mathrm{H}}
\def \emr{\mathcal{E}_\mathrm{MR}}
\def \ptot{\mathrm{P_{tot}}}
\def \snr{\mathsf{SNR}}
\def \los{\mathrm{LoS}}
\def \nlos{\mathrm{NLoS}}

\begin{document}

\title{ 
{UAVs over mmWave/THz Cellular MEC Networks: \\ A Comparative Study for Energy Efficiency}  
\author{M. Mahdi Azari, Symeon Chatzinotas
\\
}
\thanks{This work was supported by the Luxembourg National Research Fund (FNR)-5G-Sky Project, ref. FNR/C19/IS/13713801/5G-Sky, and the SMC funding program through the Micro5G and IRANATA projects.}}

\maketitle
\begin{abstract}
Cellular networks equipped with mobile edge computing (MEC) servers can be beneficial for unmanned aerial vehicles (UAVs) with limited onboard computation power and battery life-time. In this paper, we compare energy consumption of a UAV connected to cellular MEC servers in various possible scenarios such as onboard/MEC processing or parallel computation. 
Using detailed 3GPP-based modelings, we provide quantitative understanding of the most energy efficient approach and its relation with communication technologies, computation factors, and mobility parameters. Our findings show that, across the different frequencies from sub-6GHz to THz bands, the mmWave cellular MEC network is more energy efficient than UAV's onborad processing for a broader range of network densities. Secondly, while the UAV propulsion power consumption is non-decreasing function of velocity, the UAV movement cost yet can be optimized to further provide remarkable energy savings as compared to hovering. Finally, our results show that the most energy efficient approach can be obtained if mobility of UAV is combined with efficient parallel onboard-MEC processing.    
\end{abstract}


\section{Introduction}
There is growing interest in 5G and the upcoming 6G wireless networks to move the computing capabilities closer to mobile devices by locating mobile edge computing (MEC)\footnote{MEC has been referred to as mobile edge computing, but European telecommunications standards institute (ETSI) has revised it as multi-access edge computing to broaden its applications.} in the core or BSs \cite{azari2021evolution}. True convergence of communications and computing prospected in 6G empowers end-users' various devices to perform computationally expensive tasks remotely through the seamless utilization of the network computing power \cite{azari2021evolution,samsung6g}. In this manner, not only the overall latency experienced by end-users' equipment is reduced but also the devices are capable of performing computationally expensive tasks in a more energy efficient way to increase their battery life-time. 

A particular case of such devices is unmanned aerial vehicles (UAVs) which have limited energy budget and payload capacity in terms of size and weight for carrying (powerful) computing devices. Accordingly, they may not be able to perform complex computing tasks onboard and hence may rely on MEC servers for their tasks accomplishment. Increasing the number of sensors per UAV for integrated services in 6G (e.g. sensing, localization, mapping, etc.) further increases the computation tasks and the need for MEC servers \cite{azari2022thz}. However, in order for UAVs to benefit from such service, a sufficiently high-throughput communication link should be established between the UAV and the MEC network. Accordingly, the MEC service is strongly influenced by the communication network. Furthermore, proper delegation of computing tasks to MEC servers can finally result in improved UAVs battery life-time and efficient deployments. 

Considering the communication challenges of cellular-connected UAVs \cite{azari2019cellular} and significance of payload weight, computation power, and mobility impact on the overall energy consumption \cite{li2020energy}, it is of high importance to explore energy efficiency of using cellular MEC networks for UAVs. In particular, one may ask what is the most energy efficient approach between:
\begin{itemize}
    \item cellular MEC-assisted computing and UAV onboard computing? In other words, when the MEC computing could be more energy efficient? 
    \item hovering and mobility? To what extent, if applicable, UAVs movement may reduce the overall energy consumption?
    \item mixture of onboard-MEC processing and individual approach? Could it be more efficient if a mixture of onboard-MEC processing is adopted?
\end{itemize}

The integration of MEC concept and UAV networks have been considered in several prior reports, e.g. \cite{yu2020joint,tun2020energy,li2020energy,lv2020energy,yang2019energy,ei2022energy}, where the focus is mainly on the use of UAVs as MEC servers to assist the computation demand from the ground nodes. In these studies, several approaches from resource allocation and trajectory optimization to tasks scheduling are proposed without investigating the impact of 5G/6G communication technologies. However, to our knowledge, the above-mentioned questions are not yet explicitly addressed in the literature in such a way that one can appreciate the trade-offs between various communication technologies, computation alternatives, and mobility parameters. In particular, it is of high relevance to address to what extent a cellular MEC network may enhance the UAV battery-lifetime and how the energy savings can be influenced by the UAV movement, computation power, and 5G/6G cellular-related parameters. 

In this paper, we consider the energy aspect of a UAV connected to cellular network with MEC servers at the ground base stations (BSs). The goal is to address the above-mentioned questions, and how the communication technologies and computation efficiency impact the results. For this, we provide detailed 3GPP-based modelings and introduce an easy-to-follow guideline regarding the calculation of overall energy consumption for various scenarios. We propose a movement strategy based on which a fair comparison with hovering case is performed. Then, using extensive simulations, we provide insights into the energy efficiency of the scenarios. We investigate the dependency of the answers to various system parameters such as computation capacity and complexity, size of data, and the communication technology. Our key finding is that mmWave cellular MEC network almost always is more energy efficient than onboard processing while the best approach is to combine onboard-MEC computation with the UAV movement. 


\section{System Definition} \label{sec:system}

In the following we describe the network and specify the assumptions.

\subsection{Network Topology}
We consider a UAV flying at altitude $\hu$ with raw data of Q bits. 
To process the data the UAV might use an available ground cellular MEC server or an onboard computer. The UAV's onboard computer is of weight $\mcp$\,Kg and its CPU frequency is $\fcp$ cycles/s. Clearly, if the UAV does not carry any onboard computer the corresponding weight is zero, i.e. $\mcp = 0$. Furthermore, we assume that the UAV weight without the onboard computer is denoted by $\mathrm{m_0}$, and hence the UAV total weight is $\mUAV = \mathrm{m_0} + \mcp$.

To model the ground cellular MEC servers, we consider a set of ground BSs deployed on a random Poisson layout with fixed density of $\lamc$. We further assume that some of the BSs are equipped with local MEC servers to provide edge service to the users with heavy computation tasks including the UAV. Assuming that each BS is empowered and available for the MEC service with probability of $\mathrm{p_a}$, the available MEC BSs distribution follows Poisson model with density of $\lammec = \mathrm{p_a} \lamc$. The average 2D (ground) distance of the closest MEC-empowered ground BS to the UAV with uniformly 2D random location is denoted by $\mathrm{R_0}$ while the height of BS is denoted by $\hbs$. Using properties of Poisson point process one can write $\mathrm{R_0} = 1/(2\sqrt{\lammec})$ \cite{azari2020uav}.

In this work, we consider a range of possible communication technologies between UAV and BSs which represent 4G LTE, 5G mmWave, and potential 6G THz spectrum. In order to avoid remarkable LoS interference between UAV and the existing ground users in sub-6GHz frequencies the UAV-BS communication link employs orthogonal spectrum without overlap with the spectrum used by the surrounding ground users. However, such assumption for higher frequencies is less strict thanks to beamforming and higher attenuation corresponding to propagation and molecular absorption. Accordingly, the UAV-BS communication link is noise-limited. The frequency and bandwidth of communication is respectively denoted by $\fc$ and BW. 


\subsection{Propagation Channel}
We consider a dual-slop LoS/NLoS propagation channels. Each propagation channel comprises 3GPP-based large-scale path-loss fading and 3D BSs antenna gain as described in the sequel\footnote{To our knowledge, there is no 3GPP model for the THz frequency and hence we apply the widely used model presented in the literature.}.

\subsubsection{Path Loss} The path loss model used for sub-6GHz, mmWave, and THz bands are described below.

\textit{Sub-6GHz} --  The LoS and NLoS path-losses between the UAV and BS can be respectively written as \cite{3GPP36777}
\begin{align}
\mathrm{PL_\los} &= 28+22\log_{10}(\mathrm{d})+20\log_{10}(\fc), 
\end{align}
\begin{align} \nonumber
\mathrm{PL_\nlos} &= -17.5+[46-7\log_{10}(\mathrm{h_U})]\log_{10}(\mathrm{d_{3D}}) \\ &+20\log_{10}\left(\frac{40\pi\fc}{3}\right),
\end{align}
where $\mathrm{d_{3D}}$ is the 3D distance between the UAV and the BS in meters, and $\fc$ is the working frequency in GHz \cite{3GPP36777}. Note that path-loss expressions are valid for $22.5\,\mathrm{m}<\hu<300\,\mathrm{m}$ which is the range of interest for cellular-connected UAVs operations.

\textit{mmWave} -- 
The mmWave LoS path-loss can be expressed as \cite{3GPP38901}
\begin{align} \nonumber
&\mathrm{PL_\los} = 20\log_{10}\left(\frac{40\pi\mathrm{d_{3D}}\fc}{3}\right) \\ \nonumber &+ \min\left(0.03\mathrm{\hbuild^{1.72}},10\right)\log_{10}(\mathrm{d_{3D}}) \\ &- \min(0.044\hbuild^{1.72},14.77) + 0.002\log_{10}(\hbuild)  \mathrm{d_{3D}}
\end{align}
where $\fc$ is in GHz, $\mathrm{d_{3D}}$ is in meters, and $\hbuild$ is the average building heights in the range of 5\,m to 50\,m. Furthermore, the mmWave NLoS path-loss is \cite{3GPP38901}
    $\mathrm{PL_\nlos} = \max\{\mathrm{PL_\los},\mathrm{\tilde{PL}_\nlos}\}$,
where
\begin{align} \nonumber
    &\mathrm{\tilde{PL}_\nlos} = 161.04 - 7.1\log_{10}(\mathrm{W}) + 7.5\log_{10}(\hbuild) \\ \nonumber
    &- \Big(24.37-3.7(\hbuild/\hu)^2\Big)\log_{10}(\hu) \\ \nonumber
    &+ (43.42 -3.1\log_{10}(\hu))(\log_{10}(\mathrm{d})-3) \\ 
    &+ 20\log_{10}(\fc) - \Big(3.2(\log_{10}(11.75\hbs))^2-4.97\Big).
\end{align}
Above, W is the average street width between 5\! m and 50\! m.

\textit{THz} -- The path loss in THz encompasses an additional term to account for molecular absorption loss which is high and non-negligible as opposed to lower frequencies. On the other hand, due to significant loss in such frequency when facing blockages, the received NLoS signal can be ignored in many cases. Accordingly, the THz path-loss is represented only by the LoS component as
    $\mathrm{PL_\los} = \mathrm{L_P}(\fc,\mathrm{d}) + \mathrm{L_A}(\fc,\mathrm{d})$
where $\mathrm{L_P}(\fc,\mathrm{d})$ is the free space propagation loss and $\mathrm{L_A}(\fc,\mathrm{d})$ is the molecular absorption loss. Following the Beer-Lambert law, one can write
    $\mathrm{L_A}(\fc,\mathrm{d_\mathrm{U,B}}) = 4.34 \kappa(\fc) \mathrm{d_{U,B}} ~~\mathrm{[dB]}$.

Above, $\kappa(\fc)$ is the frequency-dependent medium absorption coefficient which also depends on the density and type of each gas constituent. To model the coefficient $\kappa(\fc)$, we focus on the range of 275\! GHz to 400\! GHz as the potential range of frequency in the future THz communication networks. This range of THz frequencies includes wide-band with relatively low absorption coefficients. Given such range of frequencies and following the model in \cite{kokkoniemi2018simplified}, we write
\begin{equation}
    \kappa(\fc) = \kappa_1(\fc) + \kappa_2(\fc) + \kappa_3(\fc),
\end{equation}
where
\begin{align}
    \kappa_1(\fc) &= \frac{0.2205\mu \cdot (0.1303\mu+0.0294)}{(0.4093\mu+0.0925)^2+\left(\frac{\fc}{100c}-10.835\right)^2}, \\
    \kappa_2(\fc) &= \frac{2.014\mu \cdot (0.1702\mu+0.0303)}{(0.537\mu+0.0956)^2+\left(\frac{\fc}{100c}-12.664\right)^2},
\end{align}
and $\mu$ is the volume mixing ratio of water vapor that is
calculated based on the relative humidity $\phi$ as follows
\begin{align}
    \mu &= \frac{\phi  \mathrm{p^*_\omega(T,p)}}{100\mathrm{p}}.
\end{align}
Above, $\phi \mathrm{p^*_\omega(T,p)}/100$ is the partial pressure of water vapor for which the saturated water vapor partial pressure $\mathrm{p^*_\omega}$ under pressure p and temperature T can be estimated by Buck equation as \cite{kokkoniemi2018simplified}
\begin{align}
    \mathrm{p^*_\omega(T,p)} &= 6.1121(1.0007+3.46 \times 10^{-6}\mathrm{p})e^{\frac{17.502T}{(240.94+T)}}.
\end{align}
Furthermore, $\kappa_3(\fc)$ can be written as
    $\kappa_3(\fc) = 5.54 \times 10^{-37} \fc^3 - 3.94 \times 10^{-25} \fc^2 + 9.06 \times 10^{-14} \fc -  6.36 \times 10^{-3}$.


\subsubsection{LoS Probability}
The aforementioned LoS and NLoS path-loss components are incorporated to the system along with their probability of occurrence. For $22.5\,\mathrm{m}<\hu<100\,\mathrm{m}$, the probability of LoS can be written as \cite{3GPP36777}
\[   
\mathrm{Pr_\los} = 
     \begin{cases}
       1, &\quad\text{if r} \le \mathrm{r_1}, \\
       \frac{\mathrm{r_1}}{\text{r}} + \left(1-\frac{\mathrm{r_1}}{\text{r}}\right) e^{\frac{-\text{r}}{\mathrm{r_2}}} , &\quad \text{if r} > \mathrm{r_1} \\ 
     \end{cases}
\]
where r is the 2D distance between UAV and BS, and
\begin{align}
    \mathrm{r_1} &= \max\Big(460\log_{10}(\hu)-700,18\Big), \\
    \mathrm{r_2} &= 4300\log_{10}(\hu)-3800.
\end{align}
Furthermore, for $\hu \ge 100\,\mathrm{m}$ we have $\mathrm{Pr_\los} = 1$. Finally, the NLoS probability is  $\mathrm{Pr_\nlos} = 1-\mathrm{Pr_\los}$. 

\subsubsection{Antenna Gain}
We assume that the BSs perform beamforming and are equipped with planar arrays where M and N elements are placed along the local x-axis and y-axis, respectively. The gain of an M$\times$N planar array antenna in the direction of local $(\theta,\phi)$ can be written as
\begin{equation}
    \mathrm{G(\theta,\phi) = G_E(\theta,\phi) + G_{A}(\theta,\phi)},
\end{equation}
where $\mathrm{G_E}(\theta,\phi)$ and $\mathrm{G_A}(\theta,\phi)$ are the element and array gains, respectively. 

Each element has directivity of $\mathrm{A_E}(\theta,\phi)$, where $\theta$ and $\phi$ are the spherical angles in local coordinate system of the origin at the antenna location. Following \cite{3GPP36873}, the element gain can be written as
\begin{equation}
\mathrm{A_E}(\theta,\phi) = -\min\big\{-[\mathrm{A_{E,V}}(\theta)+\mathrm{A_{E,H}}(\phi)],\mathrm{A_{m}}\big\},
\end{equation}
where 
  $  \mathrm{A_{E,V}}(\theta) = -\min\left\{12\left(\frac{\theta-90^o}{\theta_\mathrm{3dB}}\right)^2,\mathrm{SLA_v}\right\}$,
with $\theta_\mathrm{3dB} = 65^o,~\mathrm{SLA_v} = 30\,\mathrm{dB}$, and
\begin{equation}
  \mathrm{A_{E,H}}(\theta) = -\min\left\{12\left(\frac{\phi}{\phi_\mathrm{3dB}}\right)^2,\mathrm{A_m}\right\}.  
\end{equation}    
Above, $\phi_\mathrm{3dB} = 65^o,~\mathrm{A_m} = 30\,\mathrm{dB}$.
The maximum directional gain of an antenna element is considered to be $\mathrm{G_E^{max}}$. Therefore, each element's gain is 
 $\mathrm{G_E(\theta,\phi)} = \mathrm{G_E^{max}} + \mathrm{A_E}(\theta,\phi)$.

The array gain is obtained as follows
\begin{equation}
    \mathrm{G_A(\theta,\phi)} = \frac{4\pi |AF|^2}{\int_0^{2\pi}\int_0^\pi |AF|^2 \sin(\theta) \mathrm{d}\theta \mathrm{d} \phi},
\end{equation}
where the array factor (AF) is \cite[eq. 6-88]{balanis2015antenna}
\begin{align}
    \mathrm{AF(\theta,\phi)} &=  \frac{\sin(M\psi_x/2)}{M\sin(\psi_x/2)} \cdot \frac{\sin(N\psi_y/2)}{N\sin(\psi_y/2)},
\end{align}
with
   $ \psi_x = k d_x\sin(\theta)\cos(\phi) + \beta_x $ and $    \psi_y = k d_y\sin(\theta)\sin(\phi) + \beta_y $ .

Above, $d_x$ and $\beta_x$ are the spacing and progressive phase shift between the elements along the x-axis, respectively. Similarly, $d_y$ and $\beta_y$ correspond to the y-axis. Furthermore, $k = 2\pi/\lambda$ is the wave number where $\lambda$ is the wavelength. In this work, we assume $d_x = d_y = \lambda/2$. 

Finally, in order to account for the beam-forming mismatch due to for example UAV micro/macro motions we assume a deviation angle around the boresight direction which follows a normal distribution with zero mean and standard deviation of $\sigma$, i.e. $\mathcal{N}(0,\sigma^2)$. 

\subsection{Power Consumption}

The total power consumption of a UAV with communication and onboard computation capabilities can be written as
\begin{equation} \label{total_power}
    \mathrm{P_{tot}} = \mathrm{P_{cm}} + \mathrm{P_{cp}} + \mathrm{P_{pr}},
\end{equation}
where $\mathrm{P_{tot}}$ is the total power consumption, and $\mathrm{P_{pr}}$, $\mathrm{P_{cm}}$, and $\mathrm{P_{cp}}$ are respectively the power consumption of flight propulsion,  communication, and onboard computation. 
In the following, we describe each term in \eqref{total_power} individually.  

\subsubsection{Propulsion}
Several models have been proposed for the propulsion energy consumption of a rotary-wing UAV. In this work, we adopt a recent model which is based on the voltage and current flows of the electric motors  \cite{li20213d}. As opposed to the widely utilized models presented in \cite{zeng2019energy,sallouha2018energy} which are only based on the consumed energy for motion, the energy conversion efficiency of the motors is also captured in \cite{li20213d}. Therefore, following \cite{li20213d} the UAV propulsion power consumption can be written as
\begin{align}
    \mathrm{P_{pr}} &= 4\left(c_4 \omega_c^4 + c_3 \omega_c^3 + c_2 \omega_c^2 + c_1 \omega_c + c_0\right); \\
    \omega_c &= \sqrt{\frac{\mUAV \, g}{4C_T}} \left(1+\frac{C_d^2}{\mUAV^2 g^2}V^4\right)^{1/4}
\end{align}
where $V$ is the constant velocity of the UAV, $\omega_c$ is the angular speed of the motor, $\mUAV$ is the total UAV mass, and $g$ is the acceleration of gravity. Other parameters are constants defined in \cite{li20213d}.

\subsubsection{Computation}

Several works have adopted a cubic model to estimate the power consumption of computing unit as  \cite{mao2017survey} 
\begin{equation}
    \mathrm{P_{CPU}} = \eta \mathrm{f_{CPU}}^3 
\end{equation}
where it is assumed that the CPU architecture adopts the advanced dynamic frequency and voltage scaling (DVFS), $\mathrm{f_{cp}}$ is the CPU clock speed in cycle frequency and $\eta$ is the effective capacitance coefficient in $\mathrm{W/(cycle/s)^3}$ that depends on the chip architecture. However \cite{da2020characterizing} shows that in many cases with I/O intensive workflow execution the I/O operations considerably influence the overall power consumption, and hence $\mathrm{P_{cp}}$ should take into account both the I/O and CPU usage as
\begin{equation}
    \mathrm{P_{cp}} = \mathrm{P_{CPU}} + \mathrm{P_{I/O}}
\end{equation}
where $\mathrm{P_{I/O}}$ is the power consumption due to I/O operations. Other hardware components may contribute into $\mathrm{P_{cp}}$ which is ignored in compared to afore-mentioned terms \cite{mao2017survey}.



\subsubsection{Communication}
The payload communication power consumption, i.e. $\mathrm{P_{cm}}$, is negligible as compared to $\mathrm{P_{pr}}$ and hence ignored in this work \cite{zeng2019energy}.

\section{Energy Consumption Evaluation} \label{sec:cases}

In this section, we describe the scenarios and how to calculate their total energy consumption. 

\subsection{Hovering (H)} \label{sec:hovering}

In this scenario we assume that the UAV is hovering at a fixed location while performing fully onboard processing (Case A), or offloading the Q data bits to the ground MEC server for remote processing (Case B), or adopting a mixture of offloading (remote processing) and onboard processing (Case C). Assuming that the achievable throughput of the UAV-BS communication link is $\rcomm$\,bps and the onboard computation capability is $\rcomp$\,bps, then the overall time for processing of Q bits in hovering is obtained as $\th = Q/\rtot$, where
\begin{equation}
    \rtot = \rcomm + \rcomp.
\end{equation}
Note that we assume the ground MEC server computation power is high enough to neglect the processing time of the offloaded data. Using $\th$, the corresponding energy consumption is 
    $\eh = \ptot\vert_{\Vc = 0} \cdot \th$.

\begin{remark}
To study the hovering in case B, i.e. offloading all the Q bits, we simply replace $\rcomp = 0$ in the above equations. Similarly, for the case of fully onboard processing the term of $\rcomm$ is replaced with zero and hence $\rtot = \rcomp$.
\end{remark}

\subsection{Move and Return (MR)}
In this scenario the UAV moves towards the closest available MEC server with the speed of $\Vc$ and return to the initial position while transmitting the data. The idea is to increase the achievable throughput $\rcomm$ by shortening the communication distance and increasing the chance of LoS so that the overall offloading time decreases. The shorter offloading time in turn leads to lower UAV propulsion energy consumption if the mobility energy cost is not too high as compared to the hovering. In this scenario, if the UAV reaches the closest possible distance to the BS, denoted by $\mathrm{R_{min}}$, while still $Q/2$ data is not transmitted, it hovers at the minimum distance for some time and then starts to move back to the initial position to make sure the movement is energy efficient. Indeed, if the UAV moves back earlier it has to transmit the remaining data at the initial point which takes typically longer time compared to closer points. The MR scenario only is meaningful for case B and case C defined earlier in Section \ref{sec:hovering}.

In order to compute the total energy consumption of the MR, we can write
\begin{equation} \label{eq:tmr-Q}
    Q/2 = \int_0^{\tmr/2} \rtot~ \mathrm{dt},
\end{equation}
where $\tmr$ is the total time of task completion in the MR scenario. This equation relates $\tmr$ and Q. We note that at time t the 2D distance of the UAV to BS is 
\begin{equation} \label{eq:r-t}
    \mathrm{r} = \max(\mathrm{R_0-\Vc t},\mathrm{R_{min}}).
\end{equation}
After obtaining the $\tmr$ from the above equation, the energy consumption yields as 
\begin{equation} \label{eq:emr}
    \emr = \ptot\vert_\Vc \cdot \mathrm{T_{M}} + \ptot\vert_{\Vc=0} \cdot \mathrm{T_H}
\end{equation} 
where $\mathrm{T_M}$ is the time of mobility with the speed of $\Vc$ while $\mathrm{T_H}$ is the hovering time at the minimum distance to BS. 
We can write $\mathrm{T_H} = \tmr - \mathrm{T_M}$ and 
\begin{equation} \label{eq:tm-tmr}
\mathrm{T_M} = \begin{cases}
\tmr &\text{if ~$R_0 - \Vc \tmr/2 > R_{min}$}, \\
2\frac{R_0-R_{min}}{\Vc} &\text{otherwise}.
\end{cases}
\end{equation}

In the following we describe how to obtain the communication and computation rate.

The achievable communication rate (throughput) denoted by $\rcomm$, is the highest bit rate that a UAV could obtain from the network which can be written as \cite{azari2019cellular}
\begin{align} \nonumber
    \rcomm &= \mathrm{BW}~ \mathbb{E} \left[\log_2(1+\snr)\right] \\  \nonumber
    &=\mathrm{BW}~ \mathbb{E} \left[ \log_2(1+\snr_\los)\right] \mathrm{Pr_\los}  \\ \label{eq:Rcm}
    &+ \mathrm{BW}~ \mathbb{E} \left[ \log_2(1+\snr_\nlos) \right]  \mathrm{Pr_\nlos}
\end{align} 
where
\begin{equation} \label{eq:snr}
    \snr_\nu = \frac{\Ptx \mathrm{G_{tot}}/\mathrm{PL_\nu}}{ \mathrm{N_{JN} + N_m}}; ~~ \nu \in \{\los,\nlos\}.
\end{equation}
In \eqref{eq:snr}, $\mathrm{G_{tot}}$ is the total antenna gain between transmitter (UAV) and receiver (BS) nodes. Furthermore, $\mathrm{N_{JN}}$ and $\mathrm{N_m}$ are the Johnson-Nyquist and molecular noise, respectively, which can be written as
\begin{align}\label{eq:noiseJN}
    \mathrm{N_{JN}} &= \mathrm{BW} \frac{\mathrm{h_P \fc}}{\exp(\frac{\mathrm{h_P} \fc}{\mathrm{K_B T}})-1}, \\ \label{eq:noiseMol}
    \mathrm{N_m} &= \frac{\Ptx \mathrm{G_{tot}}}{\mathrm{L_P}}(1-\mathrm{L_A}^{-1}),
\end{align}
where $\mathrm{h_P}$ is Planck's constant, $\mathrm{K_B}$ is Boltzmann's constant, and $\mathrm{T}$ is the temperature in Kelvin. 

\begin{remark}
The JN noise, i.e. $\mathrm{N_{JN}}$, changes when moving towards the THz frequencies. The power of this noise remains flat at -174 \! dBm/Hz up until 100 GHz, and then decreases non-linearly where its power becomes zero at around 6 THz. Furthermore, the molecular noise, i.e. $\mathrm{N_m}$, is negligible in the considered mmWave and Sub-6GHz bands. 
\end{remark}

Finally, the computation rate of the CPU in bps can be obtained using the following
\begin{equation} \label{eq:Rcp}
    \rcomp = \frac{\fcp}{\mathrm{C_{cp}}},
\end{equation}
where $\mathrm{C_{cp}}$ represents the number of CPU cycles required to process one bit of data.

\begin{remark}
The proposed method for computing the total energy consumption of the MR scenario is summarized in the following: 1) given $\Vc,~\mathrm{R_0},~\mathrm{R_{min}}$ calculate $\rtot$ at distance r from \eqref{eq:r-t}, \eqref{eq:Rcm}, \eqref{eq:Rcp}, 2) obtain $\tmr$ from equation \eqref{eq:tmr-Q}, 3) obtain $\mathrm{T_M}$ from \eqref{eq:tm-tmr},
4) calculate $\mathrm{T_H} = \tmr - \mathrm{T_M}$, 5) calculate $\emr$ using \eqref{eq:emr}.
\end{remark}

\begin{table}[h!]
	\centering
	\caption{Notations, definitions, and default values.} \label{tab:parameters}
	\begin{tabular}{|c|c|c|}
		\hline\hline
		Notations & Definition & Default Values \\ \hline\hline
		$\hu$ & UAV altitude & 30\,m \\ \hline
		$\hbs$ & BS height & 25\,m \\ \hline
		$\hbuild$ & average building heights & 10\,m \\ \hline
		$\mathrm{W}$ & average streets width & 15\,m \\ \hline
		$\mathrm{R_0}$ & initial 2D distance (average) & -- \\ \hline
		$\mathrm{R_{min}}$ & minimum possible 2D distance & 10\,m \\ \hline
		$\fc$ & communication frequency & 2, 30, 350\,GHz \\ \hline
		BW & communication bandwidth & 1, 100, 1000\,MHz \\ \hline
		$\mathrm{d_{3D}}$ & UAV-BS 3D distance & -- \\ \hline
		$\fcp$ & onboard CPU frequency & 4\,GHz \\ \hline
		$\eta$ & effective capacitance coefficient & $10^{-28}$ \\ \hline
		$\mathrm{C_{cp}}$ & required cycles to process 1bit & 500\,cycles/b \\ \hline
		$\mathrm{m_0}$ & UAV weight without computer & 3\,Kg \\ \hline
		$\mcp$ & onboard computer weight & 500\,gr \\ \hline
		$\mUAV$ & UAV weight with computer & 3.5\,Kg \\ \hline
		Q & Total onboard data bits & 2\,Gb \\ \hline
		$\Vc$ & UAV velocity & 10\,m/s \\ \hline
		$\mathrm{P_{pr}}$ & propulsion power consumption & -- \\ \hline
		$\mathrm{P_{cp}}$ & computation power consumption& -- \\ \hline
		$\mathrm{P_{tot}}$ & UAV total power consumption & -- \\ \hline
		$\mathrm{T}$ & temperature & 300\,K \\ \hline
		$\th$ & hovering time & -- \\ \hline
		$\mathrm{T_M}$ & mobility time & -- \\ \hline
		$\tmr$ & total mobility and hovering time & -- \\ \hline
		$\mathrm{p}$ & pressure & 101325\,Pa \\ \hline
		$\mathrm{M \times N}$ & number of planar array elements & $8\times8$ mmWave  \\ 
		 & & $16\times16$ THz  \\ \hline
		 $\sigma$ & beam-forming mismatch parameter & --  \\ \hline
		$\lamc$ & BSs density & $2\times10^{-7}$ \\ \hline
		\hline
	\end{tabular}
\end{table}


\begin{figure}
\centering
\includegraphics[height=0.6\columnwidth]{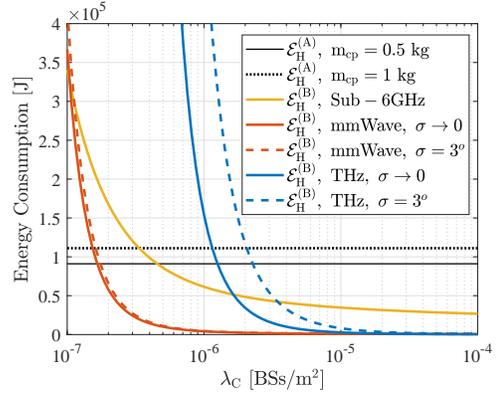}
\caption{Offloading versus onboard computing for different communication technologies.}
\label{fig:onboard_offload_lamC}
\end{figure}

\begin{figure}
\centering
\includegraphics[height=0.6\columnwidth]{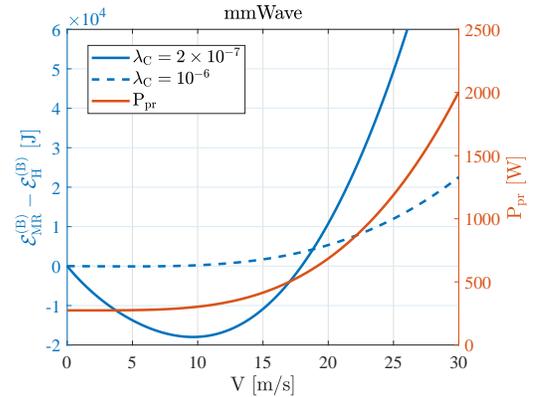}
\caption{Hovering versus mobility.}
\label{fig:hovering_mobility_lamC}
\vspace{-5mm}
\end{figure}

\section{Simulation Results} \label{sec:results}

In this section we provide the simulation results for the scenarios presented before. The default values of simulation parameters are indicated in Table \ref{tab:parameters}.

\subsection{Offloading vs. Onboard Processing}
Figure \ref{fig:onboard_offload_lamC} shows the energy consumption of hovering UAV for two cases of onboard processing (A) and offloading to MEC server (B). This figure aims to provide insights on the efficiency of MEC. In general, it is seen that the MEC service is more energy efficient only when the network is sufficiently dense. Such density of the network $\lamc$ depends on the operating frequency, and it can be seen that the lowest required density belongs to mmWave which is around $2\times10^{-7}$. When the mmWave network density is higher than this, the MEC service is significantly more energy efficient while such gain is much lower for the sub-6GHz communication. Clearly, for sparse sub-6GHz networks an onboard processing is more energy efficient. When we move to THz bands, one can see that the impact of beamforming mismatch ($\sigma = 3^o$) is significant due to much smaller beamwidth which makes the onboard processing outperform the MEC processing for low to medium densities. Comparing different communication technologies, we can see that the mmWave technology outperforms the other two frequency bands.

\subsection{Hovering vs. Mobility}

\begin{figure}
\centering
\includegraphics[height=0.6\columnwidth]{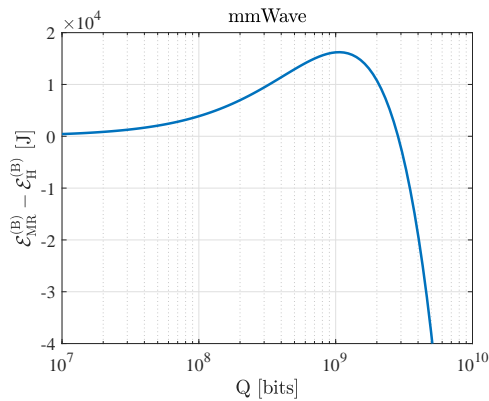}
\caption{Impact of data size at $V = 20$\,m/s.}
\label{fig:hovering_mobility_Q_mmWave}
\end{figure}

To comprehend the impact of mobility, Figure \ref{fig:hovering_mobility_lamC} compares the energy consumption of mobile scenario (MR) with that of hovering (H) when MEC service is on demand. While the propulsion power consumption is a non-decreasing function of velocity, this figure shows that the overall energy consumption with optimal mobility can be equal or less than hovering depending on network density. The velocity of $V = 10$\,m/s is where the MR has the best performance. Assuming a high velocity such as $V = 20$\,m/s where the hovering is more energy efficient, Figure \ref{fig:hovering_mobility_Q_mmWave} is obtained. As can be seen, for the large sizes of data (Q) the MR scenario is the best option. These figures show that although high velocity requires high propulsion power consumption, still \textit{movement is advantages if the size of data is big}.


\begin{figure}
\centering
\includegraphics[height=0.6\columnwidth]{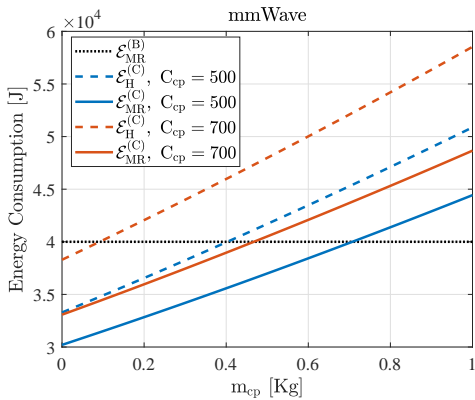}
\caption{Combination of MEC and onboard computing.}
\vspace{-5mm}
\label{fig:Mixture_mProc_Ccp}
\end{figure}

\subsection{Mixture of Onboard and MEC Processing}
Here we have the same setup as before, except that there is an onboard computer which is able to in parallel process the data (case C). Figure \ref{fig:Mixture_mProc_Ccp} compares the energy consumption of the parallel processing (case C) with the best choice of previous cases, i.e. MR without onboard computer. Interestingly, the MR with parallel processing capability, i.e. onboard-plus-MEC processing, is even more energy efficient than the benchmark, i.e. $\emr^{\mathrm{(B)}}$ for some range of $\mcp$. However, this conclusion is significantly influenced by $\mathrm{C_{cp}}$ such that less efficient algorithms and onboard CPUs power, indeed, should be lighter (lower $\mcp$) to perform more efficiently. Finally, one can observe that the parallel processing while hovering may not be more efficient than the benchmark for reasonable values of $\mcp$.

\section{Conclusion} \label{sec:conclusion}
We have studied various trade-offs in the design of energy efficient cellular MEC connected UAVs. Specifically our comparative study provided important insights regarding the impact of communication, computation, and movement parameters for an ultimate goal of energy efficiency. We concluded that UAV optimal movement while performing parallel onboard-MEC computing can be the most energy efficient approach. 

\bibliographystyle{IEEEtran}
\bibliography{Bib}

\end{document}